\documentclass{article}
\usepackage[utf8]{inputenc}
\usepackage[square,sort,comma,numbers]{natbib}
\usepackage{graphicx}
\usepackage{amsmath}
\usepackage{color}

\usepackage{geometry}
\title{Fragmentation in turbulence by small eddies}
\author{Yinghe Qi$^1$, Shiyong Tan$^1$, Noah Corbitt$^1$, Carl Urbanik$^1$, \\Ashwanth K. R. Salibindla$^1$, Rui Ni$^1$\footnote{To whom correspondence should be addressed. E-mail: rui.ni@jhu.edu}}
\date{$^1$Department of Mechanical Engineering, Johns Hopkins University, Baltimore, MD-21218, USA}

\begin{document}
\maketitle

\begin{abstract}

From air-sea gas exchange, oil pollution, to bioreactors, the ubiquitous fragmentation of bubbles/drops in turbulence has been modelled by relying on the classical Kolmogorov-Hinze paradigm since the 1950s. This framework hypothesizes that bubbles/drops are broken solely by eddies of the same size, even though turbulence is well known for its wide spectrum of scales. Here, by designing an experiment that can physically and cleanly disentangle eddies of various sizes, we report the experimental evidence to challenge this hypothesis and show bubbles are preferentially broken by the sub-bubble-scale eddies. Our work also highlights that fragmentation cannot be quantified solely by the stress criterion or the Weber number; The competition between different time scales is equally important. Instead of being elongated slowly and persistently by flows at their own scales, bubbles are fragmented in turbulence by small eddies via a burst of intense local deformation within a short time.




\end{abstract}


%



\section{Introduction}
Perhaps no other area of fluid dynamics has borne a twin problem more than bubble breakup \cite{kolmogorov1949breakage} and turbulence cascade \cite{kolmogorov1941local} both by Andrey N. Kolmogorov, based on a key idea of elementary entities, i.e. bubbles and eddies, being fragmented into smaller and smaller sizes, following a universal mechanism. In 1955, Hinze \cite{hinze1955fundamentals} extended Kolmogorov's original idea \cite{kolmogorov1949breakage}, and this Kolmogorov-Hinze (KH) framework has since posed deep and lasting impacts on modelling turbulent bubble/drop fragmentation in various flow configurations \cite{riviere2021sub, rosti2019droplets,liu2021two} and applications, including emulsion \cite{gupta2016nanoemulsions}, spray formation \cite{faeth1995structure}, and raindrop dynamics \cite{villermaux2009single}.

The key hypothesis in the KH framework is that, in turbulence, bubbles/drops with diameter $D$ are broken by eddies of the same size and the contribution from sub-bubble scale eddies is negligible. The most important dimensionless number is thus the Weber number based on $D$. The fundamental challenge associated with this key hypothesis is not about its correctness but its falsifiability. For fully-developed turbulence, eddies of many length scales are present at the same time. In these situations, bubbles always encounter eddies of various sizes, so it is extremely difficult to disentangle them cleanly \cite{cardesa2017turbulent}, not to mention establishing their roles in bubble breakup. Therefore, there has been no direct experimental evidence so far to either support or refute this hypothesis. 
To overcome the aforementioned problem and to directly test this hypothesis, we seek a ``magic knife'' to cleanly separate the sub-bubble-scale eddies from the bubble-sized ones, and expose bubbles to only one type at a time. A flow configuration that features two identical vortex rings colliding symmetrically head-on was identified as an excellent candidate because it possesses two unique stages. In the early stage, both rings expand but remain intact (Fig. \ref{fig:setup}a, the red ring), and the vortical structure only contains one large scale. As it progresses, two rings become unstable, and the large-scale flow breaks down into a turbulent cloud, comprised mainly of small eddies (Fig. \ref{fig:setup}a, the blue ring). Recently, it has been shown that the generated turbulent cloud in this later stage shares similar statistics with other types of fully-developed turbulence \cite{McKeown2018,mckeown2020turbulence}. The power spectrum follows the same -5/3 power law in the inertial range, and it lasts for an extended period of time before turbulence starts to decay. In this flow, a bubble tends to experience either only the bubble-sized vortices or turbulence consisting of the sub-bubble-scale eddies. This allows us to distinguish contributions from eddies with various sizes and directly examine the key hypothesis in the classical KH framework.

\color{black}
\section{Results}
\subsection{Turbulence production}\label{sec:cascade}

\begin{figure}[h]
    \centering
    \includegraphics[width=\linewidth]{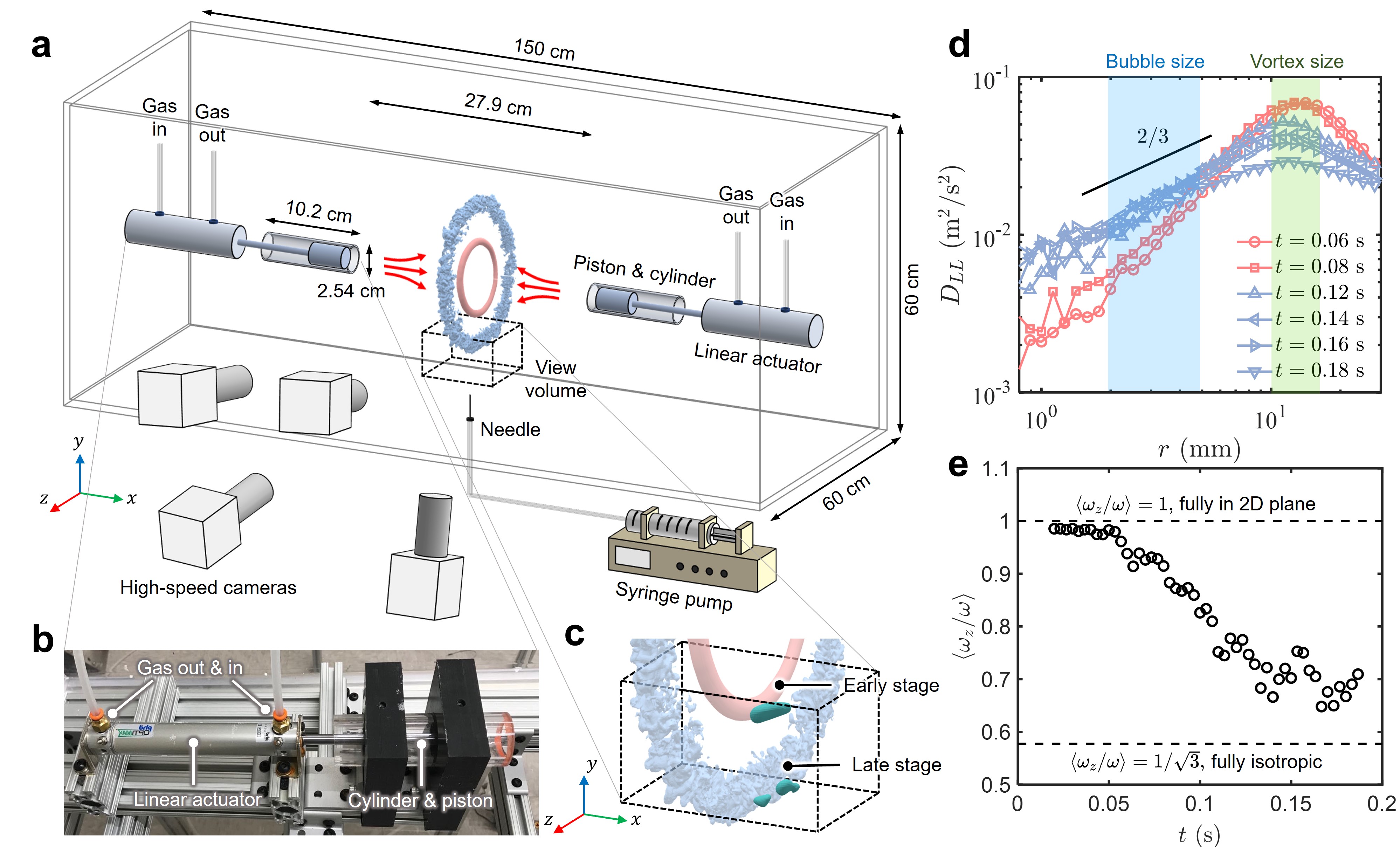}
    \caption{Experimental setup and vortex ring cascade. (a) Schematic of the experimental setup with dimensions. The black dashed box marks the location of the view volume. The smooth red ring and rough blue ring illustrate the intact vortex ring at the early stage and a turbulent cloud at the late stage, respectively; (b) Picture of the linear actuator and piston-cylinder assembly for generating the vortex rings; (c) Breaking bubbles (3D green blobs) at two different stages; (d) The longitudinal structure function $D_{LL}$; The blue and green shaded areas represent the range of the bubble size and vortex size, respectively; (e) The averaged ratio between the $z$-component vorticity $\omega_z$ and the total vorticity magnitude $\omega$ in the vortex ring as a function of time $t$. The two dashed lines indicate the two extreme limits from the uni-directional vorticity in the early stage to the fully-isotropic case later.}
    \label{fig:setup}
\end{figure}

Fig. \ref{fig:setup}a shows a schematic of the experimental apparatus that features a vortex collision sub-system (Fig. \ref{fig:setup}b) and a bubble injection sub-system. The dashed box indicates the measurement volume close to the bottom of the rings. Additional details can be found in Methods. Two distinct stages of the developed flows are highlighted in red and blue colors. The early stage was dominated by smooth and intact vortex rings, and the later stage was filled with many small eddies. Careful system control was designed to ensure that a bubble always rises to the same height when the two rings just touch each other. As shown in Fig. \ref{fig:setup}c, bubbles (indicated by the green blobs) that got entrained into one of the vortex rings were carried downward and experienced two different types of flows.

To quantify the statistics of the flow structures, the longitudinal second-order structure functions $D_{LL}$ (for details, see Supplementary Information) at different times are shown in Fig. \ref{fig:setup}d. $D_{LL}$ indicates the turbulence kinetic energy distributed at different length scales. At early times ($t<0.1$ s) after the ring collision, $D_{LL}$ exhibits a clear peak at around 10-15 mm, close to the vortex size. Most of the kinetic energy is kept within this scale. This peak decays over time as the two vortex rings become unstable and break. As the kinetic energy cascades down to smaller scales, $D_{LL}$ at these scales rise up until $t=0.12$ s when $D_{LL}$ reaches a 2/3 scaling law. This inertial-range scaling law is well known in the fully-developed turbulence based on the Kolmogorov theory \cite{kolmogorov1941local}, i.e. $D_{LL}(r)=C_2(\langle\epsilon\rangle r)^{2/3}$, which is indicated by solid lines in Fig. \ref{fig:setup}d. The prefactor $C_2\approx 2$ is the Kolmogorov constant, and $\langle\epsilon\rangle$ is the mean turbulence energy dissipation rate. From $t=0.12$ s to 0.18 s, although the kinetic energy at the vortex size continues to drop rapidly, it does not decay as much for the inertial-range scales where $D_{LL}$ exhibits the 2/3 scaling. It implies that the kinetic energy generated via the breakdown of large vortices compensates the energy dissipated at the small scales, leaving a much slower energy decay for the intermediate inertial-range scales. This inertial-range scaling lasts for roughly one integral timescale (56 ms),
which is much longer than the typical bubble breakup time that will be introduced later (See Fig. \ref{fig:modes}a bottom). This 2/3 scaling law also covers the range of bubble sizes used in the current experiments, marked by the blue shaded area. From this scaling law, the range of turbulence energy dissipation rate $\langle\epsilon\rangle$ can be estimated to be roughly 0.20-0.25 m$^2$/s$^3$.




One may expect that, as the vortex rings break down to a turbulent cloud, the flow should become more isotropic. To quantify the flow isotropy, the ratio between the $z$-component vorticity $\omega_z$ and the total vorticity magnitude $\omega$ (Supplementary Information) is shown in Fig. \ref{fig:setup}e. Two dashed lines mark the two limits of $\langle\omega_z/\omega\rangle$: $\langle\omega_z/\omega\rangle=1$ if the original vortex rings remain intact and $\langle\omega_z/\omega\rangle=1/\sqrt{3}$ if the flow becomes fully isotropic. In Fig. \ref{fig:setup}e, $\langle\omega_z/\omega\rangle$ drops gradually with time, indicating that flow indeed approaches isotropic turbulence as the cascade process continues.


\subsection{Bubble breakup modes} \label{sec:modes}



\begin{figure}[h]
    \centering
    \includegraphics[width=0.95\linewidth]{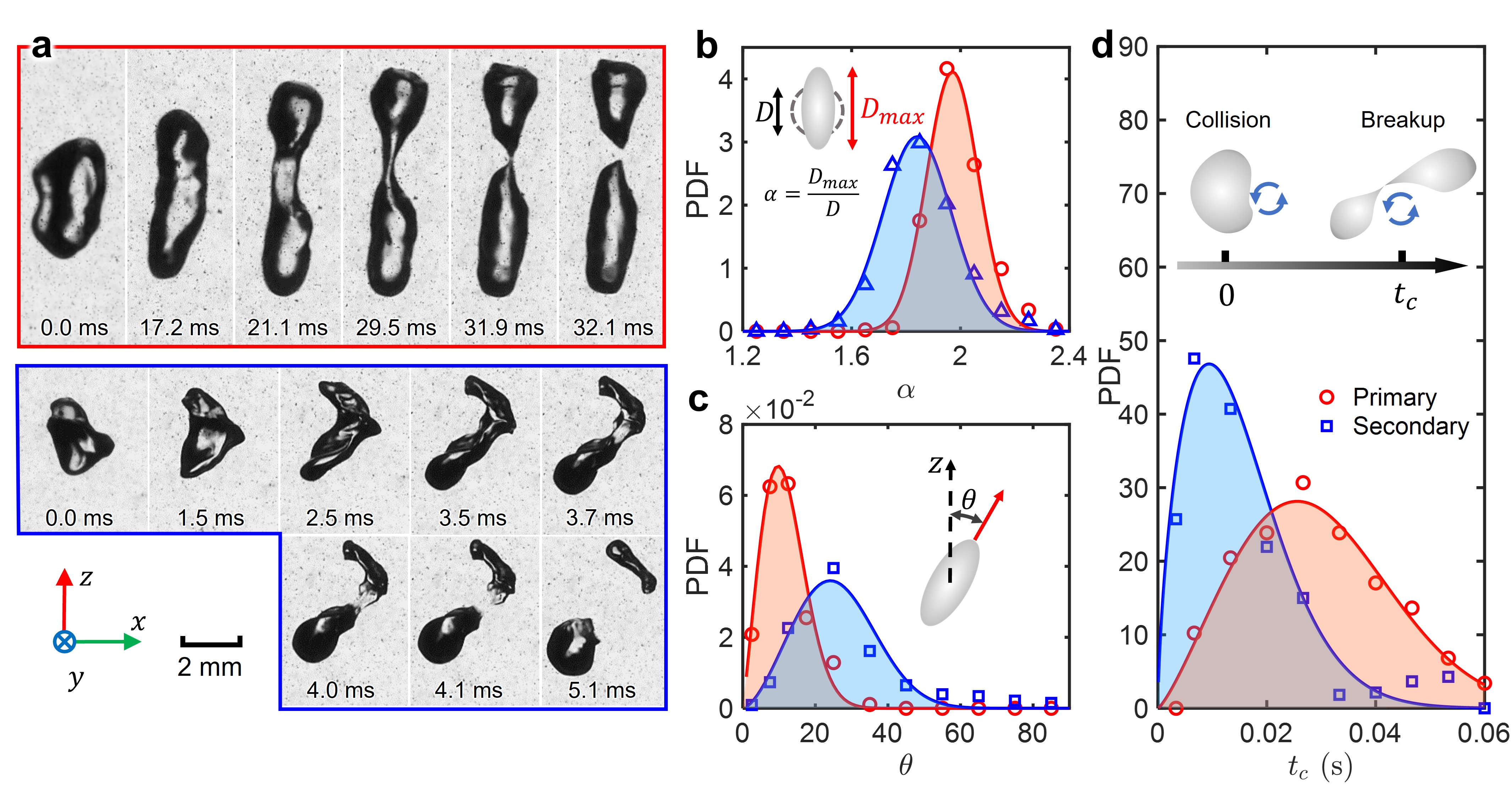}
    \caption{Geometric information of the primary and secondary breakups. (a) Examples of a primary breakup (red) driven by a smooth vortex ring and a secondary breakup (blue) driven by a turbulent cloud. More examples can be found in Supplementary Information; (b-d) The distribution of (b) the bubble aspect ratio $\alpha$, (c) the bubble orientation $\theta$, and (d) the breakup time $t_c$ for both primary (red circles) and secondary (blue squares) breakups.}
    \label{fig:modes}
\end{figure}

Once the bubble was entrained into one of the vortices at the collision point, it was carried downwards by the flow. During this process, we observed two distinct bubble breakup modes, the examples of which are shown in Fig. \ref{fig:modes}a. For the first case, a bubble was deformed consistently along the $z$-axis until the moment of breakup. This process is relatively slow, and the bubble's interface seems to be smooth throughout the entire process, similar to what was observed in the linear-extensional flows \cite{bentley1986experimental,stone1989influence}. For the rest of the paper, this type of breakup is referred to as the primary breakup as it occurs first and always before the moment when the two vortex rings break down to a turbulent cloud. 

After the primary breakup, based on the KH framework, the daughter bubbles should become harder to break because their sizes are smaller and the bubble-scale eddies have weakened, yet it is surprising to find that the daughter bubble experiences a more violent breakup, as shown in the second case of Fig. \ref{fig:modes}a. This more violent breakup is referred to as the secondary breakup hereafter. The secondary breakups have three features: (i) a rough bubble interface with large local curvatures; (ii) complicated deformation along non-persistent directions; and (iii) short breakup time. The secondary breakup occurs within 5.1 ms, which is much smaller than 32.1 ms for the primary breakup. The two breakup modes are always correlated with the bubble breakup locations. In practice, a critical height at $y_c=-51$ mm (corresponding to the vortex ring bottom location at $t=0.10$ s after their collision) was used to separate the two breakup modes (primary $y>y_c$; secondary $y<y_c$). More discussions of this separation criterion can be found in Supplementary Information.



To quantitatively compare the two breakup modes, several key statistics of the bubble geometry, orientation, and breakup time, obtained from the 3D shape reconstruction, are provided. In Fig. \ref{fig:modes}b, the probability density functions (PDFs) of the bubble aspect ratio $\alpha$, obtained from the 3D reconstructed bubble geometries from 6 ms before to the moment of breakup, for both breakup modes are illustrated. It is evident that the primary breakups typically feature a larger $\alpha$ compared with the secondary breakups. Furthermore, Fig. \ref{fig:modes}c shows the PDF of the bubble orientation, indicated by the angle between the bubble semi-major axis and the $z$-axis ($\theta$), suggesting that bubbles have preferential alignment with the $z$-axis during the primary breakup, while the distribution of $\theta$ for the secondary breakup is wider due to the disturbances from the surrounding turbulence. The third statistics that can be used to distinguish the two breakup modes is the breakup timescale $t_c$, which is defined as the time delay between the start time to the breakup instant. Note that the start time is not chosen immediately after the previous breakup, but at the minimum bubble aspect ratio closest to the breakup moment, when the bubble begins to be deformed by an eddy that will eventually break it. Fig. \ref{fig:modes}d shows the PDF of $t_c$ for the two breakup modes. The secondary breakup skews significantly more towards a smaller $t_c$ compared with the primary breakup. These three statistical quantities show a consistent picture as the two examples in Fig. \ref{fig:modes}a. 

The difference in the two breakup modes can be linked to the distinct breakup mechanisms involved. For the primary breakup, the large-scale vortex entrains the bubble towards its center---a local pressure minimum. For a bubble with a size close to the vortex diameter, as it reaches the vortex center, it experiences a pressure gradient that tends to compress it along the radial direction and extend it along the $z$-axis. Secondary breakup is more irregular, driven by a turbulent cloud filled with sub-bubble scale eddies. Bubbles break without significant elongation because the process is disrupted and accelerated by these eddies with smaller timescales, which also explains the observed difference in breakup time $t_c$.

We emphasize that the primary breakup follows the key hypothesis made in the classical KH framework, in which a bubble is assumed to be broken by a clean and isolated vortex filament with a size close to the bubble diameter. However, most bubble breakups observed in fully developed turbulence are closer to the secondary case, where the contribution from a cloud of smaller eddies cannot be ignored.

\subsection{Bubble breakup mechanism} \label{sec:mechanism}

\begin{figure}[h]
    \centering
    \includegraphics[width=0.9\linewidth]{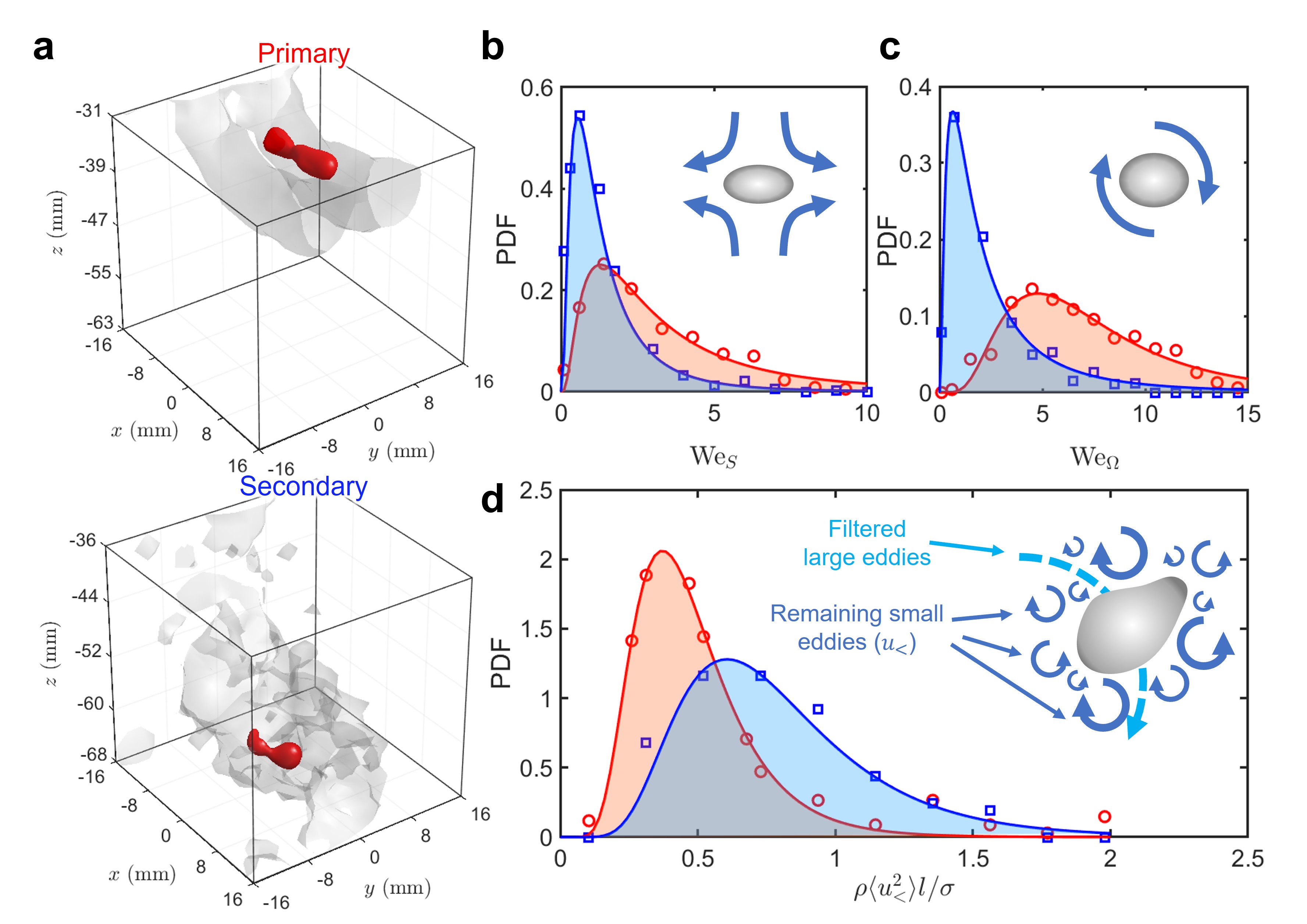}
    \caption{Visualization of flows around breaking bubbles and the PDFs of Weber numbers. (a) Experimentally reconstructed bubble geometries (red) and the isosurfaces of the surrounding vorticity for the primary and secondary breakups; (b-d) The distribution of (b) $\text{We}_S$, (c) $\text{We}_\Omega$, and (d) the non-dimensionalized sub-bubble-scale velocity variance, $\langle u_<^2\rangle$, around breaking bubbles (primary, red circles; secondary, blue squares). The solid lines represent the log-normal fits to the experiment data.}
    \label{fig:mechanism}
\end{figure}

To understand these two breakup modes, we measured the 3D flow in the vicinity of a breaking bubble along with its 3D geometry to show two examples of the primary and secondary breakups in Fig. \ref{fig:mechanism}a. In particular, the 3D isosurfaces of the vorticity magnitude at $t=0.06$ s for the primary breakup and $t=0.14$ s for the secondary breakup are used to illustrate the flow structures. For the primary breakup, the bubble is trapped within the bottom portion of one vortex ring and the flow surrounding the bubble is smooth, whereas the secondary breakup takes place in a chaotic turbulent cloud, indicated by the rough isosurface of vorticity. 

Based on the 3D flows, the first quantity that can be acquired is the Weber number. At the bubble scale, the flow can be decomposed into the straining and rotational components. In practice, the velocity gradient coarse-grained at a scale close to the bubble size, i.e. $\widetilde{A}_{ij}$, is extracted based on the velocity of multiple tracer particles around the bubble (for details, see Supplementary Information). From $\widetilde{A}_{ij}$, the coarse-grained strain rate tensor $\widetilde{S}_{ij}$ and rotation tensor $\widetilde{\Omega}_{ij}$ can be determined: $\widetilde{S}_{ij}=(\widetilde{A}_{ij}+\widetilde{A}_{ji})/2$ and $\widetilde{\Omega}_{ij}=(\widetilde{A}_{ij}-\widetilde{A}_{ji})/2$, which lead to the definition of two Weber numbers:
\begin{equation}\label{eqn:weber_strain_rot}
    \text{We}_{S}=\frac{\rho(|\lambda_3| D)^2 D}{\sigma}, \quad \text{We}_{\Omega}=\frac{\rho(\omega D)^2 D}{\sigma}
\end{equation}
where $\lambda_3$ (the largest compression rate) is the smallest eigenvalue of $\widetilde{S}_{ij}$, and $\omega$ is the vorticity magnitude. The new definition of the two Weber numbers extends the original one-dimensional version in the KH framework to emphasize the contributions from the 3D straining and rotational flows. Nevertheless, the key assumption in the KH framework that the only relevant length scale is the bubble size is still applied here.



Fig. \ref{fig:mechanism}b and \ref{fig:mechanism}c show the PDFs of $\text{We}_{S}$ and $\text{We}_{\Omega}$ over 6 ms duration before the moment of breakup for both the primary and secondary modes. For the primary breakup, $\text{We}_{\Omega}$ appears to be systematically larger than $\text{We}_{S}$ because the bubble is compressed more by the radial pressure gradient due to the flow rotation than by the straining flow. For the secondary breakup, the peaks of the PDFs of $\text{We}_{S}$ and $\text{We}_{\Omega}$ both locate at values smaller than one, and more importantly, smaller than their primary breakup counterparts.

The KH framework implies that bubbles with larger Weber numbers tend to break more easily. If it were right, we should expect a more violent primary breakup. However, the observations suggested otherwise, which clearly refute the key hypothesis in the KH framework. For the secondary breakup, although the eddy of the bubble size is much weaker, many sub-bubble-scale eddies begin to emerge. To demonstrate their appearance, we apply a high-pass rolling-average spatial filter with a filter length $l=3$ mm (which is selected to be close to the bubble mean diameter) to the velocity field. The residual fluctuation velocity $u_<$ and its variance $\langle u_<^2 \rangle$ only contain the contribution from small eddies ($D_e<l$). Fig. \ref{fig:mechanism}d shows the PDF of non-dimensionalized residue fluctuation velocity around bubbles for both primary and secondary breakups. The secondary breakup has systematically larger $\langle u_<^2 \rangle$ compared with the primary counterpart, which implies that the contribution of sub-bubble scale eddies that were missed in the KH framework may be the key to understand the bubble breakup in turbulence. 


\subsection{Model}\label{sec:model}



\begin{figure}[h]
    \centering
    \includegraphics[width=\linewidth]{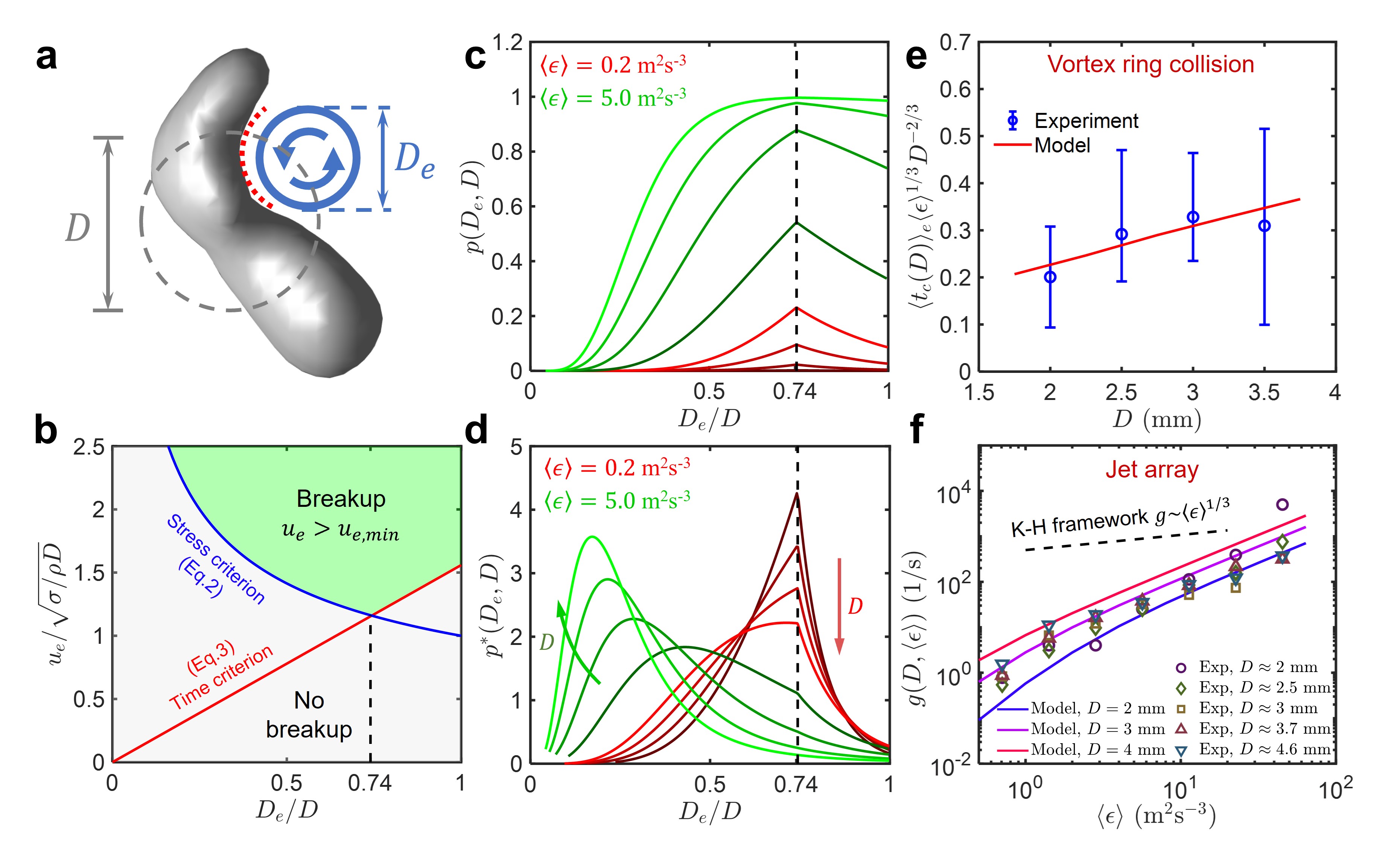}
    \caption{Bubble breakup model and its experimental validations. (a) Schematic of a bubble with an equivalent spherical diameter $D=4$ mm deformed by an eddy of size $D_e$ in turbulence; (b) The eddy velocity $u_e$ required to break a bubble versus the dimensionless eddy size $D_e/D$ based on the proposed new dimensionless numbers: We$_e$ (blue) and Ti (red), and their imposed criteria. These two criteria together split the entire plot into the breakup (green) and no breakup (grey) regions; (c) The probability $p(D_e,D)$ of a bubble (size $D$, ranging from 2 mm (dark color) to 5 mm (light color)) broken by a single eddy (size $D_e$) as a function of $D_e/D$ at two different $\langle\epsilon\rangle$=0.2 m$^2$/s$^3$ (red) and 5.0 m$^2$/s$^3$ (green). The dashed line marks the peak location at $D_e/D=0.74$; (d) The weighted breakup probability $p^*(D_e,D)$ versus $D_e/D$ (the color is consistent with that adopted in (c)); (e) The normalized expected bubble breakup time $\langle t_c(D) \rangle_e$ as a function of the bubble size $D$ for experiments (blue symbols) and model (red line). The upper and lower bounds of the error bars mark the 30th and 70th percentiles of the bubble breakup time, respectively; (f) The breakup frequency $g(D,\langle\epsilon\rangle)$ of bubbles with different sizes in turbulence with various mean energy dissipation rates $\langle\epsilon\rangle$, including the experimental data by Vajrazka et al. \cite{vejrazka2018} (symbols) and predictions from our model (solid lines). The dashed line shows the scaling of $g(D,\langle\epsilon\rangle)\propto \langle\epsilon\rangle^{1/3}$ predicted by the KH framework.}
    \label{fig:model}
\end{figure}

To extend the KH framework, we argue that both the bubble diameter and eddy size are relevant length scales. To include both, let us consider a simple scenario: a bubble with an equivalent spherical diameter of $D$ encounters and gets deformed by a small eddy of size $D_e<D$, as shown in Fig. \ref{fig:model}a. Consistent with the features observed in the secondary breakup in Fig. \ref{fig:modes}a, the concave interface has a large local curvature that is presumably linked to the eddy size $D_e$, so the interfacial stress can be determined by using $\sigma/D_e$, instead of $\sigma/D$ in the KH framework. Furthermore, the eddy inertia $\rho u_e^2$ has to overcome the interfacial stress to break the bubble. Therefore, a new Weber number based on the eddy velocity scale $u_e$ and length scale $D_e$ can be defined, and this Weber number ($\text{We}_e$ based on the eddy size $D_e$) has to be larger than one for the bubble to break,
\begin{equation}\label{eq:criteria1}
  \text{We}_e=  \frac{\rho u_e^2 D_e}{\sigma}>1~~~~~~~\text{(Stress criterion)}
\end{equation}

In addition to this stress constraint, another key missing piece in the KH framework is time. The eddy timescale has to be shorter than the bubble's relaxation time, otherwise, the bubble will return back to the sphere. The eddy lifetime or turnover time scales with $D_e/u_e$, whereas the bubble relaxation time can be estimated by using the natural frequency of the bubble, i.e. $\sqrt{96\sigma /(\rho D^3)}/(2\pi)$ (Lamb mode 2 \cite{lamb1932}). So the second criterion can be set as $D_e/u_e<2\pi\sqrt{\rho D^3/(96 \sigma)}$. Rearranging this relationship yields another new dimensionless number ($\text{Ti}$, which represents the time ratio) that involves both $D$ and $D_e$: 
\begin{equation}\label{eq:criteria2}
 \text{Ti}=   \frac{\rho u_e^2 D^3 / D_e^2}{\sigma} > \frac{96}{4\pi^2}~~~~~~~\text{(Time criterion)}
\end{equation}

Eq. \ref{eq:criteria1} and \ref{eq:criteria2} together provide two new dimensionless numbers, We$_e$ and Ti, which set two constraints on the eddy velocity $u_e$, whose dependence on the eddy size $D_e$ are shown as blue and red solid lines in Fig. \ref{fig:model}b. Based on these two constraints, the minimum eddy velocity to break a bubble ($u_{e,min}(D_e,D)$) can be estimated following:
\begin{equation}\label{eq:eddy_vel_min}
    u_{e,min}(D_e,D)=\max\left(\sqrt{\frac{\sigma}{\rho D_e}},\sqrt{\frac{96}{4\pi^2}\frac{\sigma}{\rho D^3/D_e^2}}\right)
\end{equation}

This equation divides Fig. \ref{fig:model}b into two regions: breakup ($u_e>u_{e,min}(D_e,D)$) and no breakup ($u_e\leq u_{e,min}(D_e,D)$). It is evident that $u_{e,min}$ has a non-monotonic dependence on $D_e/D$, reaching its own minimum at the intersection of the two lines: $\sqrt{\sigma/(\rho D_e)}=\sqrt{96 \sigma/(4\pi^2 \rho D^3/D_e^2)}$, which results in $D_e/D=0.74$.



To estimate the breakup probability, we ignore the bubble-induced flow modulation and assume the surrounding turbulence follows the same statistics as their single-phase counterpart, the distribution of the eddy velocity can be calculated based on the log-normal distribution of the energy dissipation rate \cite{meneveau1991,kolmogorov1962} follows:
\begin{equation}\label{eq:epsilon_pdf}
\begin{split}
    P(\epsilon_e)=&\frac{1}{\epsilon_e}\frac{1}{\sqrt{2\pi\sigma_{\ln\epsilon}^2}}\exp{\left[-\frac{\left(\ln{(\epsilon_e/\langle\epsilon\rangle)}+\sigma_{\ln\epsilon}^2/2\right)^2}{2\sigma_{\ln\epsilon}^2}\right]}
\end{split}
\end{equation}
where $\sigma_{\ln\epsilon}^2=A+\mu\ln(L/D_e)$ is the variance; $A$ represents a large-scale variability, which is set at $A=0$ for convenience; $\mu\approx0.25$ is the intermittency exponent; and $L$ is the integral length scale of turbulence. In this work, the original vortex diameter is used to estimate the integral scale, which is $L\approx15$ mm; we use $\langle\epsilon\rangle$ = 0.20 m$^2$/s$^3$ based on the previous estimation. The instantaneous eddy velocity $u_e$ can be directly related to the eddy-size based dissipation rate, following $u_e=\sqrt{2}(\epsilon_e D_e)^{1/3}$. As a result, the PDF of the eddy velocity, $P(u_e|D_e)$, for any given eddy size $D_e$ can be expressed as
\begin{equation}\label{eq:eddy_vel_pdf}
    P(u_e|D_e)=\frac{3\sqrt{2}}{2}\epsilon_e^{2/3} D_e^{-1/3}P(\epsilon_e)
\end{equation}

Based on this eddy velocity distribution, the probability of a bubble of size $D$ being broken by a single eddy of size $D_e$, $p(D_e,D)$, can be calculated by following $p(D_e,D)=\int_{u_{e,min}}^\infty P(u_e|D_e)du_e$, where eddies with a velocity larger than $u_{e,min}$ (Eq. \ref{eq:eddy_vel_min}) are integrated. Fig. \ref{fig:model}c (the red-curve group) shows $p(D_e,D)$ versus the non-dimensionalized eddy size $D_e/D$ for $\langle\epsilon\rangle$=0.2 m$^2$/s$^3$. In contrast to the hypothesis made in the KH framework, a wide range of eddies smaller than the bubble size can drive bubble breakup; in fact, eddies with $D_e$ close to 0.74$D$ are actually much more efficient in breaking a bubble than bubble-sized eddies. Furthermore, $p(D_e,D)$ of all $D_e$ increases systematically as $D$ grows.

So far, the discussions were limited to the breakup probability driven by a single eddy without considering that smaller eddies are more abundant. To account for this effect, the collision rate $\omega_c$ (for details, see Supplementary Information) is used to define a weighted breakup probability ($p^*(D_e,D)$) following the expression of: $p^*(D_e,D)=\omega_c p(D_e,D)/\int_{10\eta}^D \omega_c p(D_e,D) d D_e$, which is shown in Fig. \ref{fig:model}d as red curves for different bubble sizes at $\langle\epsilon\rangle$=0.2 m$^2$/s$^3$. It appears that bubbles of all sizes are still preferentially broken by eddies of size $D_e=0.74D$. But the contributions by even smaller eddies are growing as $D$ increases.

In addition to the size dependence, $p(D_e,D)$ and $p^*(D_e,D)$ for the same range of $D$ but using a higher $\langle\epsilon\rangle$ at 5.0 m$^2$/s$^3$ is shown as green curves in Fig. \ref{fig:model}c and d, respectively. In Fig. \ref{fig:model}c, as $\langle\epsilon\rangle$ increases, the probability ($p(D_e,D)$) of one bubble being broken by one eddy of size $D_e$ increases for all $D_e$, while maintaining its peak at $D_e=0.74D$, until $p(D_e,D)$ saturates for a range of $D_e$. But once accounting for the number density difference of eddies of different sizes, the peak locations of $p^*(D_e,D)$ for all bubble sizes shift to a much smaller $D_e$, suggesting that bubbles are preferably fragmented by much smaller eddies in stronger turbulence. This result implies that the deviation from the hypothesis in the KH framework becomes more significant for either larger bubbles or stronger turbulence.

\color{black}


To validate the proposed model, we quantify the breakup time of the secondary breakups since these events are driven by small eddies. The expected bubble breakup time for different bubble sizes $D$ is shown as open circles in Fig. \ref{fig:model}e, and their respective distributions are indicated by the error bars with the upper and lower bounds marking the 30th and 70th percentiles, respectively. 

In the model, if we assume that the breakup time scales with $D/u_e$, the mean breakup time ($t_c(D_e,D)$) for a bubble of size $D$ being broken by an eddy of size $D_e$ can thus be expressed as: $t_c(D_e,D)=\int_{u_{e,min}}^\infty D/u_e P(u_e|D_e)du_e/\int_{u_{e,min}}^\infty P(u_e|D_e)du_e$. By considering the contribution of all sub-bubble scale eddies within the inertial range ($10\eta<D_e<D$), the expected bubble breakup time can finally be estimated following:
\begin{equation}\label{eq:breakup_time}
   \langle t_c(D) \rangle_e=\frac{\int_{10\eta}^D t_c(D_e,D)\omega_c dD_e}{\int_{10\eta}^D \omega_c dD_e}
\end{equation}
where $\langle...\rangle_e$ represents the integration over all eddy sizes. The model predicted breakup time $\langle t_c(D) \rangle_e$ is shown in Fig. \ref{fig:model}e in comparison with the experimental results of the secondary breakups. A nice overall agreement can be observed without using any fitting parameters. 





To ensure that the generalization of the model to other types of turbulent flow configuration is appropriate, in addition to our experiments, we also compare the model prediction against another dataset in fully-developed  turbulence driven by a jet array \cite{vejrazka2018}. In this experiment, the breakup frequency $g(D,\langle\epsilon\rangle)$ is measured for different bubble sizes ($D$) over a wide range of $\langle\epsilon\rangle$. The results are shown in Fig. \ref{fig:model}f as open symbols. The dashed line indicates the scaling predicted based on the KH framework by assuming the breakup frequency scales with the reciprocal of the turnover time of the bubble-sized eddies, i.e. $(\epsilon D)^{1/3}/D\propto \epsilon^{1/3}$. It is evident that the measured breakup frequency exhibits a much steeper scaling than the 1/3 power law suggested by the KH framework.



Based on our model, $g(D,\langle\epsilon\rangle)$ can be estimated by including the contributions from all small eddies, following $g(D,\langle\epsilon\rangle)=\int_{10\eta}^D \omega_c p(D_e,D) d D_e$. The model predictions for different $\langle\epsilon\rangle$ and $D$ are shown in Fig. \ref{fig:model}f as solid lines. It can be seen that the modeled breakup frequency agrees with the experimental results well, capturing the steeper scaling that the data suggests. This scaling originates from the growing contributions from small eddies with a higher frequency as $\langle\epsilon\rangle$ increases. 
Furthermore, our model suggests that there is no simple power law relationship between the breakup frequency and $\langle\epsilon\rangle$ because of the non-trivial dependence of $p^*(D_e,D)$ on $D$ and $\langle\epsilon\rangle$.




\color{black}

\section{Discussion}\label{sec:conclusion}

The classical KH framework hypothesized that bubbles tend to be broken by eddies of similar sizes in turbulence. So the only length scale used in the definition of the Weber number is the bubble diameter ($D$), and the contributions by the sub-bubble-scale eddies are ignored.  To directly test this hypothesis, one would need to conceive an experiment that can cleanly disentangle eddies of different sizes. To that end, we developed a experiment by injecting a bubble into the flow driven by the head-on collision between the two vortex rings. As the rings break down into turbulence, the flow experiences two distinct stages with the early stage consisting of only the large vortical structures of the bubble size and the late fully-developed turbulence filled with a spectrum of the sub-bubble-scale eddies. In this flow, two distinct bubble breakup modes were observed. Against our intuition, the more violent and rapid secondary breakup occurred when the bubble size was smaller, the flow kinetic energy was lower, and the resulting Weber number was much weaker. This observation clearly shows the inadequacies of employing only the Weber number and the bubble length scale, disapproving the key hypothesis in the KH framework. A model for bubble breakup in turbulence based on two new dimensionless numbers that account for two scales, the bubble and eddy sizes, is therefore proposed. Our framework supplement the classical framework that relies only on the stress criterion by adding a key criterion--the time scale. Our model is validated by the excellent agreement between the model prediction and experiments from two seemingly different flow configurations. Our work emphasizes the importance of the sub-bubble scale eddies for bubble fragmentation and adds a dimension to the existing framework on the bubble/drop dynamics in turbulence \cite{villermaux2007fragmentation, mathai2020bubbly, lohse2018bubble}.


\section{Methods}
\paragraph{Vortex ring}

Vortex rings were generated using two identical piston-cylinder assemblies in a water tank with a size of 60$\times$60$\times$150 cm$^3$. The cylinders and pistons were carefully adjusted to make sure they are aligned. Both cylinders have an inner diameter $D_0$ of 2.54 cm with the separation between their exits about 27.9 cm. Within each cylinder, a piston is driven by a stainless steel shaft connected to a pneumatic linear actuator. The piston can be moved forward and backward at different stroke times $T_s$ controlled by the air pressure. The stroke length $L_s$ was kept at 10.2 cm, and the stroke ratio was, therefore, $SR=L_s/D_0=4$. Each assembly can generate vortex rings with different Reynolds numbers $Re=V_p D_0/\nu$ based on the mean velocity of the piston $V_p=L_s/T_s$ and the kinematic viscosity of the water $\nu$. A hypodermic needle with a 1.7 mm inner diameter connected to a syringe filled with air was used to generate bubbles. By altering the injection rate through a computer-controlled syringe pump and needle size, bubbles with diameters ranging roughly from 2 to 5 mm can be generated. In this study, the initial bubble size is fixed at around 5 mm.

\color{black}

\paragraph{Time delays and system control}

One critical element of our experiments is the timing. Bubbles must be entrained into the vortices exactly at the moment when the two rings collide with each other. Therefore, three sub-systems, including two vortex-ring generators, one syringe pump for bubble injection, and four cameras were all controlled by a digital data acquisition (DAQ) system. Since $t=0$ s is the reference time when two vortex rings began to touch each other, the syringe pump was activated at $t_1=-1.34$ s for bubbles to naturally rise into the collision point; the vortex-ring generators were actuated at $t_2=-0.63$ s for vortex rings to travel from the generators to the collision point; and the cameras were triggered at $t_3=-0.18$ s to capture some frames before the collision. Even though the three time delays can be accurately controlled by our DAQ system, bubbles or vortex rings may not arrive at the exact same moment from one run to another. In order to quantify the uncertainty of the arrival time, we used high-speed cameras (7500 fps with 0.13 ms uncertainty) to capture the arrival time of rings and bubbles. The measured uncertainty of the arrival time difference between rings and bubbles is about $\pm$38 frames, corresponding to about $\pm$5.1 ms. This number is much smaller than the time that it takes for the two vortex rings to break down into turbulence, i.e. about 100 ms, or the integral timescale of the turbulence generated, i.e. about 60 ms, which suggests that the synchronization is sufficiently accurate for reproducible experiments.

\paragraph{3D measurements}
In each run, the shadows of bubbles and many surrounding tracer particles were projected onto all four cameras by the back LED panels. The images of individual phases were segmented based on the contrast and size differences. For the liquid phase, the 3D tracer trajectories were determined by performing the Lagrangian particle tracking with the in-house OpenLPT code \cite{tan2020} that implemented the parallelized Shake-The-Box algorithm \cite{schanz2016shake}. These trajectories were then smoothed by convoluting them with the Gaussian kernels \cite{mordant2004experimental,ni2012lagrangian}, from which the particle velocity and acceleration could be obtained along their trajectories. For the gas phase, the bubble 3D geometry was reconstructed using the virtual-camera visual hull method \cite{masuk2019robust} that calculates the intersection of the cone-like volume extruded from the silhouette on each camera. Once the 3D bubble geometry was obtained, the bubble center of mass was linked frame by frame to get the bubble trajectory.

\paragraph{Statistics}

In total, we collected 83 datasets and observed about 150 breakup events for vortex ring Reynolds number ranging from 7.5$\times10^4$ to 1.1$\times10^5$. The majority of the data was taken at the largest Reynolds number, in which almost each experiment yielded both primary and secondary breakups. For smaller Reynolds number, the turbulence generated is not sufficiently intense to fragment bubbles. So, these datasets were not included in the statistics. The breakups that were used in this manuscript are plotted as a point cloud in the Supplmentary Fig. S7 to show the breakup locations relative to the flow.

\color{black}

\section{Data availability}
All the data supporting this work are available from the corresponding author upon reasonable request.

\section{Code availability}
The OpenLPT code used for 3D particle tracking is available at https://github.com/JHU-NI-LAB/Open
\\
LPT\_Shake-The-Box (DOI: 10.5281/zenodo.5750942). Other codes for image processing and analyses that are reported in the paper are available from the corresponding author upon request.

\bibliographystyle{plain} 
\bibliography{main}

\section{Acknowledgements}

We acknowledge the financial support from the National Science Foundation under the award numbers: 1854475, CAREER-1905103 to RN. This project was also partially supported by the ONR award: N00014-21-1-2083 to RN. Any opinions, findings, and conclusions or recommendations expressed in this material are those of the author(s) and do not necessarily reflect the views of the Office of Naval Research. The authors are also grateful for the data provided by Dr. Rodolfo Ostilla-M{\'o}nico and the useful discussion with Dr. Charles Meneveau.

\section{Author contributions}

Y.Q. and R.N. designed the research; N.C and C.U. built the experiment setup; Y.Q., S.T. and A.S. developed the 3D measurement techniques; Y.Q. conducted the experiments and analyzed the data; R.N. managed the project; Y.Q. and R.N. wrote the initial manuscript; all authors discussed the results and commented on the final manuscript.

\section{Competing interests}
The authors declare that they have no known competing financial interests or personal relationships that could have appeared to influence the work reported in this paper.

\end{document}